\pdfoutput=1
\documentclass[aps,prl,twocolumn,superscriptaddress,a4paper,floatfix]{revtex4}
\usepackage[latin1]{inputenc}
\usepackage{graphicx}
\usepackage{amsmath,amssymb}
\usepackage{hyperref}
\usepackage{datetime}
\usepackage{pgfpages}
\usepackage{tikz}

\newcommand{\supp}[1] {#1\,dB}
\def \mol {9.4(8)}
\def \cold {4.6(3)}
\def \hot {2.1(2)}

\begin{document}

\title{Interferometric Measurement of Local Spin-Fluctuations in a Quantum Gas}

\author{Jakob Meineke}
\affiliation{Institute for Quantum Electronics,  ETH Zurich, 8093
Zurich, Switzerland}
\author{Jean-Philippe Brantut}
\affiliation{Institute for Quantum Electronics,  ETH Zurich, 8093
Zurich, Switzerland}
\author{David Stadler}
\affiliation{Institute for Quantum Electronics,  ETH Zurich, 8093
Zurich, Switzerland}
\author{Torben M\"uller}
\affiliation{Institute for Quantum Electronics, ETH Zurich, 8093
Zurich, Switzerland}
\author{Henning Moritz}
\affiliation{Institut f\"ur Laser-Physik, Universit\"at Hamburg, 22761 Hamburg, Germany}
\author{Tilman Esslinger}
\email{esslinger@phys.ethz.ch}
\affiliation{Institute for Quantum
Electronics, ETH Zurich, 8093 Zurich, Switzerland}

\maketitle

{\bf The subtle interplay between quantum-statistics and interactions is at the origin of many intriguing quantum phenomena connected to superfluidity and quantum magnetism \cite{Auerbach_Quantum_Magnetism}. The controlled setting of ultracold quantum gases is well suited to study such quantum correlated systems \cite{varenna_lectures}. Current efforts are directed towards the identification of their magnetic properties \cite{jordens_quantitative_2010, Nascimbene2011, sommer_universal_2011}, as well as the creation and detection of exotic quantum phases \cite{lewenstein_ultracold_2007,eckert_quantum_2008, roscilde_quantum_2009}. In this context, it has been proposed to map the spin-polarization of the atoms to the state of a single-mode light beam \cite{bruun_probing_2009}. Here we introduce a quantum-limited interferometer realizing such an atom-light interface \cite{hammerer_quantum_2010} with high spatial resolution. We measure the probability distribution of the local spin-polarization in a trapped Fermi gas showing a reduction of spin-fluctuations by up to 4.6(3)\,dB below shot-noise in weakly interacting Fermi gases and by 9.4(8)\,dB for strong interactions. We deduce the magnetic susceptibility as a function of temperature and discuss our measurements in terms of an entanglement witness.}

Quantum mechanics manifests itself in the correlations between the constituent parts of a physical system. These correlations quantify the probability of joint measurements, and are experimentally observable in the statistical distribution of the outcomes of repeated measurements. Experiments studying density fluctuations have successfully demonstrated the potential of these techniques as a tool to study local thermodynamic properties of quantum gases \cite{esteve_observations_2006, gemelke_situ_2009, muller_local_2010, sanner_suppression_2010, bakr_probing_2010, sherson_single-atom-resolved_2010, serwane_deterministic_2011}. In another context, interferometric methods have been used to study spin-fluctuations in atomic vapors, leading to the observation of entanglement and spin-squeezing \cite{hammerer_quantum_2010, oblak_quantum-noise-limited_2005, appel_mesoscopic_2009}. More recently, several authors have proposed to apply similar techniques to map quantum fluctuations, generated by the many-body dynamics in a quantum gas \cite{eckert_quantum_2008, roscilde_quantum_2009, bruun_probing_2009}, on to the optical field of a single mode probe beam. In a different approach, speckle noise originating from out-of-focus regions in off-resonant imaging was related to the spin-fluctuations of a Fermi gas \cite{sanner_speckle_2011}. In this letter, we use a shot-noise limited interferometer to directly measure the probability distribution of the local spin-fluctuations in a two-component quantum degenerate Fermi gas.

\begin{figure}[t]
  \includegraphics[width=0.5\textwidth]{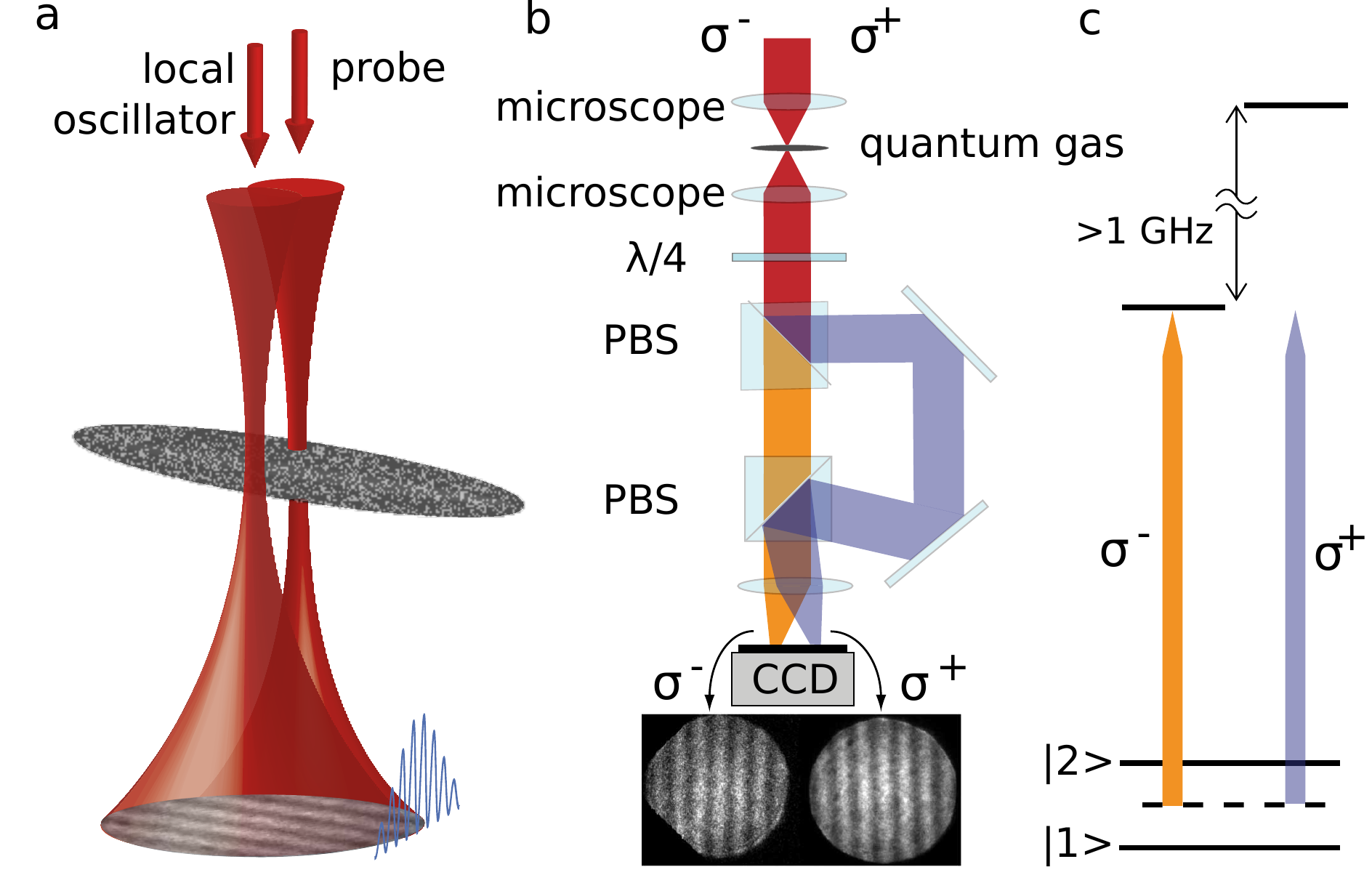}
  \caption{Interferometer Setup (a) Interferometer beams in the vicinity of the atomic cloud. While the probe passes through the cloud shown in grey, the local oscillator passes by the side of it. The beams (minimum $1/e^2$-radius $1.2\, \mu{\rm m}$) overlap in the far field giving rise to an interference pattern as shown. See methods for creation of the interferometer beams. (b) Optical setup to obtain two interference patterns, only one of which is affected by the atoms. Using a quarter-wave retardation plate ($\lambda/4$) and two polarizing beam splitters (PBS), the $\sigma^-$-component of the polarization, which interacts with the atoms, is separated from the $\sigma^+$-component. This yields two far-field interference patterns on one image as shown in the lower part of the figure, see methods. The lens adjusts the size of the patterns on the camera. (c) Level-scheme illustrating that only $\sigma^-$ light interacts with the atoms. $\sigma^+$ light is far detuned from resonance. Transitions from both states to their respective excited states are nearly closed cycling transitions.} \label{fig1}
\end{figure}

Our interferometer is analogous to Young's double slit experiment. Two tightly focused beams, the probe and the local oscillator, are focused to separate points as shown in Figure~\ref{fig1} and overlap in the far field. Position and visibility of the resulting interference pattern are determined by changes in phase and amplitude of the probe beam, which passes through an atomic cloud, while the local oscillator does not. The analysis of the interference pattern thus allows the reconstruction of both quadratures of the probe beam, phase and amplitude, which carry information about the local properties of the atomic cloud.

A probe beam passing through a mixture of $^6$Li atoms in the lowest two hyperfine states $|1\rangle$ and $|2\rangle$ acquires a phase-shift given by
\begin{equation}
  \phi=\sigma_0\left(\frac{n_1\delta_1}{1+s+4\delta_1^2}+\frac{n_2\delta_2}{1+s+4\delta_2^2}\right),
  \label{eq1}
\end{equation}
with $\sigma_0$ the resonant scattering cross-section, $s$ the saturation parameter and $n_i$ and $\delta_i$ the line-of-sight integrated density and the frequency detuning in atomic linewidths for state $|i\rangle$, respectively. By choosing the detuning exactly in between the two resonances, $\delta_1=-\delta_2=6.4$, the phase shift $\phi$ is proportional to the line-of-sight integrated spin-polarization density $m=n_1-n_2$. A single measurement of the phase-shift yields the spin-polarization of the given experimental realization. Consequently, the full probability distribution of spin-polarization can be reconstructed from repeated measurements.

\begin{figure}[t]
  \includegraphics[width=0.5\textwidth]{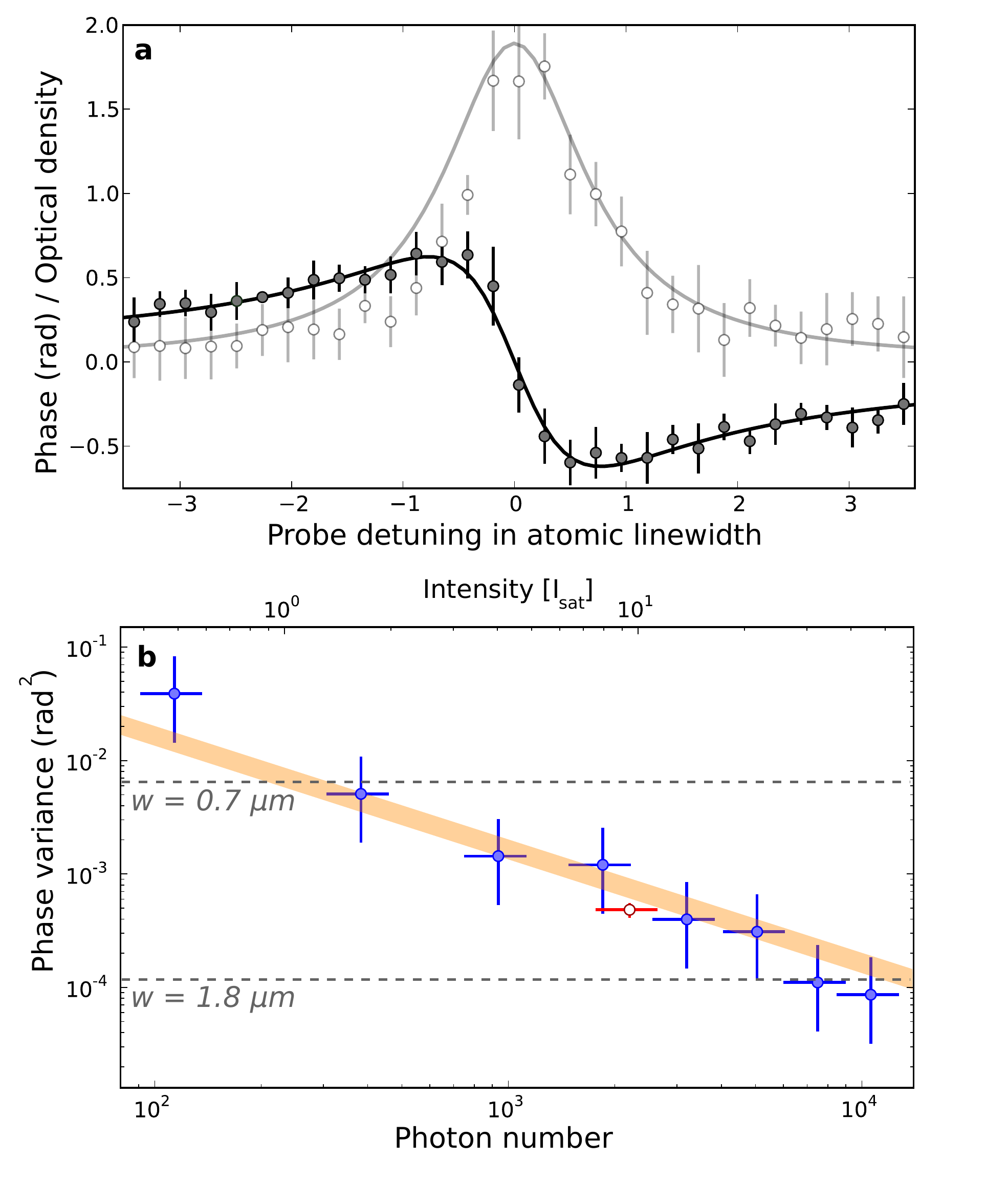}
  \caption{Interferometer Performance (a) Measured phase (\protect \tikz {\protect \filldraw[fill=black!50, draw=black] (0,0) circle (2pt);}) and optical density (\protect \tikz {\protect \filldraw[draw=black!50,fill=white] (0,0) circle (2pt);}) as a function of detuning. Maximum saturation of the probe beam $s=1.2$, duration of the probe pulses $1.2\, \mathrm{\mu s}$. Solid lines result from a fit to the phase data using the model described in the text yielding $n_2\sigma_0=3.2$, effective saturation $0.6$ and an effective linewidth that is 20\% broader than the natural linewidth of $5.9\,\mathrm{MHz}$, which we attribute to the probe laser. Errorbars show standard deviations. (b) Measured phase-variance (without atoms) $\delta \phi^2$ (\protect \tikz {\protect \filldraw[fill=blue!50, draw=blue] (0,0) circle (2pt);}) as a function of photon number in the probe beam determined from 100 measurements for each point. The empty circle (\protect \tikz {\protect \filldraw[fill=white, draw=red] (0,0) circle (2pt);}) indicates the phase-variance for the intensity at which the spin-polarization measurements were made. Errorbars are an uncorrelated sum of statistical and systematic uncertainties. Expected phase-noise (\protect \tikz {\protect \fill[red!50!yellow!50] (0pt,2pt) -- (10pt,2pt) -- (10pt, 7pt) -- (0pt,7pt) --cycle;}) for quantum efficiency $\eta=0.6$. The width of the line corresponds to 20\% errors estimated from the uncertainty of our determination of $\eta$. The gray lines indicate the square of the phase shift expected for a single atom fixed in space at a detuning of half the atomic linewidth for the indicated $1/e^2$-waists of the probe beam \cite{aljunid_phase_2009}.
}\label{fig2}
\end{figure}

To validate our procedure we show that (i) equation (\ref{eq1}) holds in the parameter regime of the experiment, and (ii) that the phase measurement is only limited by photon shot-noise of the probe beam. Then, each additionally detected photon leads to a projection of the atomic state into a smaller subspace. To verify (i), we measure the frequency-dependent phase shift and optical density for a Fermi gas comprised of atoms in state $|2\rangle$ only, see methods for preparation and data-processing. Figure~\ref{fig2}a shows the characteristic asymmetric profile of the phase whereas the optical-density is well-described by a Lorentzian. The solid lines result from fitting the phase data to equation (\ref{eq1}) with $n_1=0$ and agree with the measurement provided the probe duration is $\sim 1\,{\rm \mu s}$ and the saturation is less than $\sim 10$. The use of stronger (longer) pulses leads to a systematic shift of measured phases to larger values, due to the light-forces resulting from the strong focusing of the probe beam and scattering of photons. To verify (ii), we measure the phase-variance as a function of the photon number as shown in Figure~\ref{fig2}b. From the power-law behavior of the phase-variance we deduce that photon shot-noise limits the sensitivity of the phase measurement. Indeed, the phase-noise is expected to be given by $\delta \phi^2=\frac{1}{\eta N}$ \cite{lye_nondestructive_2003}, where $N$ is the number of photons and $\eta=0.6$ the quantum efficiency, which is determined in an independent measurement.

\begin{figure*}[t]
  \includegraphics[width=\textwidth]{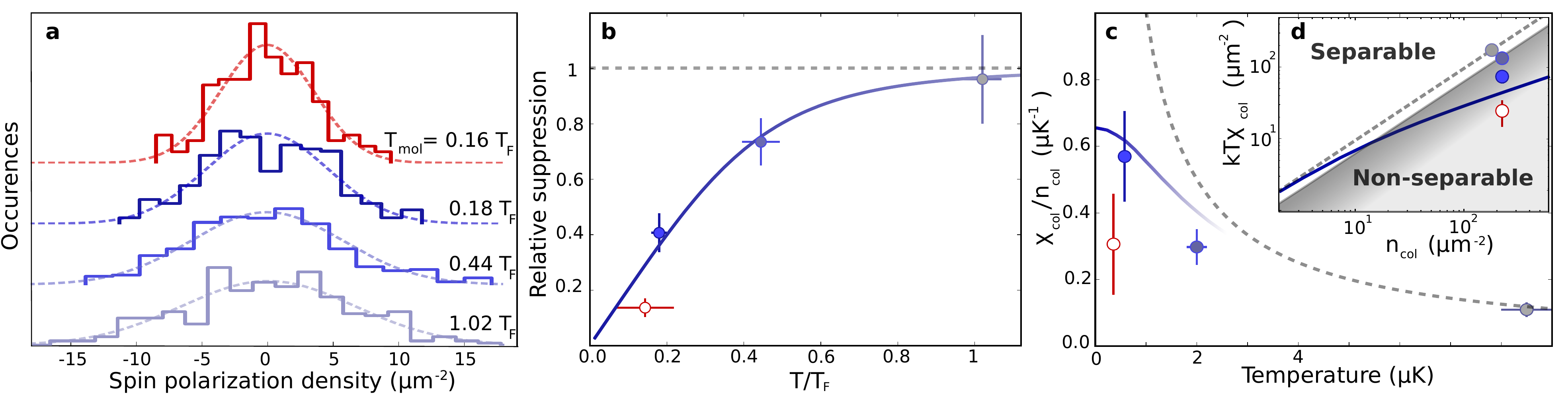}
  \caption{Spin-Fluctuations in Ultracold Fermi Gases (a) Normalized histograms for the measured spin-polarization for weakly interacting gases at $T_1=1.02(2)\,T_F$ (\protect \tikz {\protect \draw[white] (0pt,0pt) -- (10pt,0pt); \protect \draw[blue!40!white,very thick] (0pt,2pt) -- (10pt,2pt);}), $T_2=0.44(5)\,T_F$ (\protect \tikz {\protect \draw[white] (0pt,0pt) -- (10pt,0pt); \protect \draw[blue!75!white,very thick] (0pt,2pt) -- (10pt,2pt);}) and $T_3=0.18(2)\,T_F$ (\protect \tikz {\protect \draw[white] (0pt,0pt) -- (10pt,0pt); \protect \draw[blue,very thick] (0pt,2pt) -- (10pt,2pt);}) as well as for a strongly interacting gas of molecules at $T_{{\rm mol}}=0.16(7)\,T_F$ (\protect \tikz {\protect \draw[white] (0pt,0pt) -- (10pt,0pt); \protect \draw[red,very thick] (0pt,2pt) -- (10pt,2pt);}), where $T_F$ is the Fermi temperature. (b) Reduction of spin-fluctuations compared to a thermal gas of equal column density as a function of $T/T_F$. Data points are colored identical to (a). The solid line shows theory for a noninteracting Fermi gas. (c) Spin susceptibility per atom $\chi_{\rm col}/n_{\rm col}$ as a function of temperature for the weakly interacting gas (blue points) and the strongly interacting gas (\protect \tikz {\protect \filldraw[fill=white, draw=red] (0,0) circle (2pt);}). Also shown are the susceptibility for a classical gas (\protect \tikz {\protect \draw[white] (0pt,0pt) -- (10pt,0pt); \protect \draw[black!50!white,very thick, dashed] (0pt,2pt) -- (10pt,2pt);}), which varies as $1/T$ and the theoretical expectation (\protect \tikz {\protect \draw[white] (0pt,0pt) -- (10pt,0pt); \protect \draw[blue!40!white,very thick] (0pt,2pt) -- (10pt,2pt);}) for the trap parameters of the gas at $T_3$. (d) Spin-fluctuation density $A \delta m^2=kT\chi_{\rm col}$ as a function of column-density $n_{\rm col}$. The prediction for a thermal gas (\protect \tikz {\protect \draw[white] (0pt,0pt) -- (10pt,0pt); \protect \draw[black!50!white,very thick, dashed] (0pt,2pt) -- (10pt,2pt);}) is shown along with the boundary separating the separable from the non-separable region. The blue line shows theory for a noninteracting Fermi gas at temperature $T_{{\rm mol}}=0.36\,{\rm \mu K}$. Errorbars are an uncorrelated sum of statistical and systematic uncertainties.
  } \label{histo}
\end{figure*}

We first measure the distribution of the spin-polarization for weakly interacting Fermi gases at three different temperatures given by $T_1=8.5(5)\,{\rm \mu K}$, $T_2=2.0(2)\,{\rm \mu K}$ and $T_3=0.58(5)\,{\rm \mu K}$, see methods for preparation. Figure~\ref{histo}a shows that the probability distributions of the spin-polarization have a narrower width the lower the temperature of the gas. The distributions exhibit no significant asymmetry and are well described by Gaussian functions, as expected from the large number of atoms in the probe beam, $\sim\,500$ in each state. For weakly interacting Fermi gases, number fluctuations in each hyperfine state are independent, so that fluctuations of the spin-polarization are given by $\delta m^2=\delta n_1^2+\delta n_2^2$. Consequently, with the onset of quantum-degeneracy, spin-fluctuations are reduced, since in each state only atoms at the Fermi edge contribute to the fluctuations, which is a manifestation of antibunching due to Pauli's principle \cite{muller_local_2010, sanner_suppression_2010}. The measured variances of the spin-polarization, in order of decreasing temperature, are $\delta m_{T_1}^2=35(6)\,{\rm \mu m^{-4}}$, $\delta m_{T_2}^2=27(3)\,{\rm \mu m^{-4}}$ and $\delta m_{T_3}^2=15(3)\,{\rm \mu m^{-4}}$. Here, the background corresponding to a standard deviation of $\pm 13(2)$ atoms in the probed volume has been subtracted, see methods. The column density in the probed region was $n_1=n_2=n_{{\rm col}}/2=110\,{\rm \mu m^{-2}}$ for the gases at $T_2$ and $T_3$ and $20\,\%$ lower for the gas at $T_1$. From our measurement we determine a reduction of the spin-fluctuations as compared to a thermal gas prepared with the same column density by \supp{\hot} for the gas at $T_2$ and \supp{\cold} for the gas at $T_3$, as shown in Figure~\ref{histo}b, which is in quantitative agreement with theory for a noninteracting Fermi gas, see methods.

We now turn to the study of a gas with strong repulsive interactions, prepared close to a Feshbach resonance at temperature $T_{{\rm mol}}=0.36(10)\,{\rm \mu K}$, see methods. The resulting histogram is displayed in Figure~\ref{histo}a and shows a distribution of the spin-polarization that is significantly narrower than for the weakly interacting gases. This reflects the fact that for interacting gases correlations are present between the different spin states. In particular, for strong repulsive interactions close to a Feshbach resonance, weakly bound molecules form \cite{varenna_lectures} and the number of atoms in the two states will always be equal. Spin-fluctuations in the atom number difference are created at the cost of breaking molecules and are consequently suppressed. We measure $\delta m_{{\rm mol}}^2=5(2)\,{\rm \mu m^{-4}}$ at $n_{\rm col}/2=110\,{\rm \mu m^{-2}}$, with the background subtracted as before. This corresponds to a reduction by \supp{\mol} as compared to a noninteracting thermal gas. A reduction by $\sim 18 \rm{dB}$ is expected from ref. \cite{bruun_probing_2009} for a molecular BEC at zero temperature. The observation of a lower value could be caused by pair-breaking or fluctuations of the probe frequency, see methods.

The measured values for the spin-fluctuations can be used to determine the magnetic susceptibility $\chi$ via the fluctuation-dissipation theorem (FDT) \cite{seo_compressibility_2011}, provided the probed system is in grand-canonical equilibrium with its surroundings. Due to column-integration, the fluctuation-dissipation theorem here reads $kT \int\chi=kT \chi_{{\rm col}}=A \delta m^2$, where $\chi_{{\rm col}}$ is the column-integrated magnetic susceptibility, $k$ Boltzmann's constant and $A$ the effective area of the probed column. For small volumes and at low temperatures corrections to the FDT are expected, because correlations between the probed system and its surroundings cannot be neglected \cite{klawunn_local_2011, hung_observation_2011}. Using ref. \cite{klawunn_local_2011} and including column-integration, we estimate the corrections to be less than 10\% even for our coldest samples. Figure~\ref{histo}c shows the column-integrated magnetic susceptibility per particle $\chi_{{\rm col}}/n_{{\rm col}}$ as a function of temperature. It relates the spin imbalance to the energy needed to create it. For high temperatures, the susceptibility decreases inversely proportional to the temperature as expected. For low temperatures and with the onset of quantum degeneracy, the susceptibility saturates to a value depending on the trap details: the stiffer the trap, the lower the susceptibility. The solid line shows theory for noninteracting fermions calculated for the trap parameters of the gas at $T_3$. The susceptibility for the strongly interacting gas of molecules is lower than for the weakly interacting gas at $T_3$, despite its lower temperature and weaker trap.

The link between the magnetic susceptibility and spin-fluctuations has been proposed as a macroscopic entanglement witness in solid state systems \cite{wiesniak_magnetic_2005}. We now apply this concept to our measurement of the spin-fluctuations. For this purpose we describe each particle as a two-level system, which together form an effective spin $\mathbf{M}$ ~\cite{wiesniak_magnetic_2005, toth_spin_2009}, where $\langle M_z\rangle$ is proportional to the atom number difference and $\delta M_z^2=A^2 \delta m^2$. For all separable states, the ratio of the spin-fluctuations to the total atom number is bounded from below. For a system in grand-canonical equilibrium described by a Hamiltonian which is invariant under rotations of the spin state, this bound is expressed by the inequality
\begin{equation}
\frac{A \delta m^2}{n_{{\rm col}}}\geq\frac{2}{3}, \label{eq:inequality}
\end{equation}
see methods and supplementary informations for details. Figure~\ref{histo}d shows the spin-fluctuations $A \delta m^2$ as a function of the column density $n_{\rm{col}}$. The gas at $T_3$ violates the above inequality, which is expected because the Pauli principle leads to non-separable states at low temperatures~\cite{Vedral2003}. The notion of entanglement applied to indistinguishable particles is subject of current investigations~\cite{Horodecki2009}.

For the case of the strongly interacting gas, we find $\frac{A \delta m^2}{n_{{\rm col}}}=0.11(4)$, violating the bound by more than 10 standard deviations. The fluctuations are $2.4\,\rm{dB}$ lower than expected for a noninteracting gas at the same temperature, see bold line in Figure~\ref{histo}d. This corresponds to $2.4\,\rm{dB}$ of spin-squeezing following ref. \cite{kheruntsyan_quantum_2006}.

Our analysis uses the following assumptions: (i) We create a gas with equal atom number in the two states, which leads to $\langle M_z\rangle=0$. (ii) We probe the system in thermal equilibrium. As a consequence, the transverse components have dephased to the amount permitted by Pauli's principle and we have $\langle M_x \rangle=\langle M_y \rangle=0$. (iii) The time-evolution is described by a Hamiltonian that is invariant under rotations of the spin~\cite{Zwierlein2003}. It is therefore sufficient to measure only the $z$-component.

The rigorous application of the entanglement witness~\cite{wiesniak_magnetic_2005, toth_spin_2009} requires equal coupling of all atoms to the probe beam. Here, we take into account the inhomogeneity of the probe beam by introducing the effective area $A$.

In conclusion, we have measured the probability distribution of the spin-flucutations in a trapped Fermi gas. We have discussed our measurement in terms of an entanglement witness. Our work constitutes a first step towards the detection of entanglement in many-body states in quantum gases. The detection of higher order correlations could be achieved by extracting higher moments of the probability distribution for the spin-polarization~\cite{cherng_quantum_2007}.

{\bf Methods:}

\paragraph{Generation of interferometer beams} 
The interferometer beams are generated by applying two radio frequencies differing by 20~MHz along each of the axes of a two-axis acousto-optical deflector (AOD) very similar to previous work \cite{zimmermann_high-resolution_2011}. This results in four beams in the \mbox{-1/-1} diffraction order of the AOD which, after passing through a high-resolution microscope objective, are arranged in a square. Two of these beams, probe beam and local oscillator, have exactly the same frequency and form the interference pattern on the camera. The other two are detuned by $\pm20$~MHz and their interference patterns average out over the duration of the probe pulses. The intensities of the beams are controlled via the power in the individual radio frequencies so that the local oscillator is 20 times as intense as the probe beam. Phase stability is ensured by deriving each radio frequency from the same source for both axes. The light beams are elliptically polarized. Due to the birefringence of the atomic cloud in a magnetic field, also used in polarization-contrast imaging \cite{bradley_bose-einstein_1997}, only the $\sigma^-$-polarized component of the light interacts with the atoms, while the $\sigma^+$- component passes undisturbed, see Figure~\ref{fig1}b. The power ratio of $\sigma^+$- and $\sigma^-$-component is about 10.

\paragraph{Experimental sequence} An equal mixture of $^6$Li atoms in the two lowest hyperfine states, denoted by $|1\rangle$=$|m_J$=$-1/2,m_I$=$1\rangle$ and $|2\rangle$=$|m_J$=$-1/2,m_I$=$0\rangle$ is prepared similar to previous work \cite{muller_local_2010}. A second dipole trap with a wavelength of $767\, {\rm nm}$ and a $1/e^2$-radius of $10\,{\rm \mu m}$ is then switched on over 500\,ms and is used to locally increase the total column-density. Finally the magnetic field is ramped to 475\,G where the scattering length is $a=-100\,a_0$, with $a_0$ the Bohr radius. The final trap depths of the large dipole trap are $84\,{\rm \mu K}$ for $T_1$, $19\,{\rm \mu K}$ for $T_2$ and $10\,{\rm \mu K}$ for $T_3$ corresponding to $1050\, {\rm mW}$, $235\, {\rm mW}$ and $130\, {\rm mW}$, respectively. The trap depths of the second dipole trap are $9\,{\rm \mu K}$, $4.5\,{\rm \mu K}$ and $1.5\,{\rm \mu K}$, respectively. The central region of the cloud is probed interferometrically with a pulse of 1.2\,$\mu$s duration at a maximum saturation of 9. Experiments are repeated 400 times. Images of the whole cloud are taken after the interferometric measurement to determine the total atom number as well as the temperature from the radial expansion after $1\,{\rm ms}$ time of flight. Experiments that show a larger deviation of the total atom number than 5\% are discarded, amounting to $10-20\%$ of the images.
For the preparation of the strongly interacting gas, we ramp directly to $800\, {\rm G}$, where $a=7000\,a_0$ and the trap depth is lowered to $17\, {\rm mW}$ followed by recompression to $30\, {\rm mW}$, which corresponds to a trap depth of $4.8\, {\rm \mu K}$ for the molecules. The experiment is repeated 200 times. We estimate the final temperature to be one tenth of the relevant trap depth leading to $T_{{\rm mol}}=0.36(10)\,{\rm \mu K}$ after recompression. For the preparation of a gas containing only atoms in state $|2\rangle$, we hold the gas for 100\,ms close to a $p$-wave Feshbach resonance at $159\,{\rm G}$, which leads to the loss of nearly all particles in state $|1\rangle$. Subsequent evaporation of the remaining atoms leads to a non-degenerate gas of atoms in state $|2\rangle$.

\paragraph{Theory}
We compare our measurements for the weakly interacting gases with theory for noninteracting fermions \cite{Huang_Statistical}. We determine the temperature and the chemical potential from fits to the time-of-flight images as in previous work \cite{muller_local_2010}. We then calculate the density distribution in the combined trap given the fitted temperature and atom number to determine $T_F$ in the center of the combined trap. This assumes full thermalization in the presence of the second trap, which is supported by the measured spin-fluctuations. The knowledge of $T$, $\mu$ and the trap shape allows us calculate the mean and the variance of the atomic density along the line of sight using column-integration.

\paragraph{Data processing} Three images are obtained in each experiment, one with atoms in the interferometer and two without. The images are averaged along the direction parallel to the fringe-pattern and a Fourier filter is applied to suppress the low spatial frequencies. For the determination of the mean phase, a simultaneous sinusoidal fit to the $\sigma^-$-pattern on images with and without atoms is used to determine the phase. Free parameters are amplitude and wavelength of the fringe pattern, as well as a common phase for both patterns and a phase-difference for the picture with atoms. Residual mechanical motion, e.g. due to the microscopes, is compensated for by using the same simultaneous fit applied to the $\sigma^+$-pattern. For determination of the phase-fluctuations we find it advantageous to analyze the correlations between the $\sigma^+$- and the $\sigma^-$-pattern on each image. This allows us to exploit the similarity of the two patterns and use the $\sigma^+$-pattern to noiselessly amplify the signal contained in the $\sigma^-$-pattern, analogous to homodyne techniques. The phase-shift due to the atoms causes a displacement of the zero-crossings of the correlation function in the image with atoms as compared to the image without atoms. See supplemental materials for further details on data processing.

\paragraph{Determination of effective area} The effective area $A$ relates the spin-fluctuations to the mean atomic density and corresponds to the area of a beam with uniform intensity giving the same result. For the nearly non-degenerate gas at $T_1=1.02 T_F$ the spin-fluctuations are well described by Poissonian statistics because the atoms are uncorrelated. The fluctuations are thus proportional to the average atom number in the probe volume. This allows the determination of the effective area $A=0.97 n_{{\rm col}}/\delta m_{T_1}^2=4.9(8)\,{\rm \mu m^2}$, corresponding to an effective waist of the probe beam of $1.8(3)\,{\mu m}$. The factor $0.97$ accounts for the residual suppression of the spin-fluctuations at $T_1=1.02 T_F$.

\paragraph{Background} The contribution to the phase variance originating from photon shot-noise is determined from the two images without atoms using the above described procedure. This yields $\delta m_{{\rm bgr}}=6.7(1.0)\,{{\rm \mu m^{-4}}}$ and a standard deviation of the atom number difference in the probed volume of $\sqrt{A^2 \delta m_{\rm bgr}^2}=13(2)$.
Frequency fluctuations of the probe can cause fluctuations in the measured phase. A variance of the probe frequency of 2\,MHz$^2$ would correspond to apparent spin-fluctuations of $5\,{\rm \mu m^{-4}}$ at the column-density in our experiment and could contribute to the difference of our results compared to the theory in ref. \cite{bruun_probing_2009}.

\paragraph{Entanglement Witness} Each atom realizes a two-level system represented by Pauli matrices $\sigma_{x,y,z}$. We define the spin-polarization or magnetization of the probed region as $M_{x,y,z}=\sum_i \sigma_{x,y,z}^i$, where the sum extends over all atoms in the probe volume. For all separable states the inequality $\delta M_x^2+\delta M_y^2+\delta M_z^2\geq 2 \langle N\rangle$ is fulfilled, where $\langle N\rangle$ is the average number of atoms probed \cite{wiesniak_magnetic_2005}. Using the invariance of the Hamiltonian under spin rotations and that we measure expectation values in the grand-canonical ensemble, which show the same symmetry as the Hamiltonian, this reduces to $\frac{\delta M_z^2}{N}=\frac{A \delta m^2}{n_{\rm col}} \geq\frac{2}{3}$. See Supplementary Information for more details.

%

\begin{thebibliography}{10}

\bibitem{Auerbach_Quantum_Magnetism}
Auerbach, A.
\newblock {\em Interacting Electrons and Quantum Magnetism}.
\newblock Springer,  (1998).

\bibitem{varenna_lectures}
Inguscio, M., Ketterle, W., and Salomon, C., editors.
\newblock {\em Ultra-cold Fermi Gases: Proceedings of the International School
  of Physics "Enrico Fermi"}, volume Course CLXIV,  (2008).

\bibitem{jordens_quantitative_2010}
J\"ordens, R., Tarruell, L., Greif, D., Uehlinger, T., Strohmaier, N., Moritz,
  H., Esslinger, T., Leo, L.~D., Kollath, C., Georges, A., Scarola, V., Pollet,
  L., Burovski, E., Kozik, E., and Troyer, M.
\newblock Quantitative determination of temperature in the approach to magnetic
  order of ultracold fermions in an optical lattice.
\newblock {\em Phys. Rev. Lett.}{ \bf 104}(18), 180401, May  (2010).

\bibitem{Nascimbene2011}
Nascimb\`ene, S., Navon, N., Pilati, S., Chevy, F., Giorgini, S., Georges, A.,
  and Salomon, C.
\newblock {Fermi-Liquid} behavior of the normal phase of a strongly interacting
  gas of cold atoms.
\newblock {\em Physical Review Letters}{ \bf 106}(21), 215303, May  (2011).

\bibitem{sommer_universal_2011}
Sommer, A., Ku, M., Roati, G., and Zwierlein, M.~W.
\newblock Universal spin transport in a strongly interacting fermi gas.
\newblock {\em Nature}{ \bf 472}(7342), 201--204, April  (2011).

\bibitem{lewenstein_ultracold_2007}
Lewenstein, M., Sanpera, A., Ahufinger, V., Damski, B., Sen, A., and Sen, U.
\newblock Ultracold atoms in optical lattices: Mimicking condensed matter
  physics and beyond.
\newblock {\em Adv. Phys.}{ \bf 56}, 243--379 (2007).

\bibitem{eckert_quantum_2008}
Eckert, K., {Romero-Isart}, O., Rodriguez, M., Lewenstein, M., Polzik, E.~S.,
  and Sanpera, A.
\newblock Quantum non-demolition detection of strongly correlated systems.
\newblock {\em Nat. Phys.}{ \bf 4}(1), 50--54, January  (2008).

\bibitem{roscilde_quantum_2009}
Roscilde, T., Rodriguez, M., Eckert, K., {Romero-Isart}, O., Lewenstein, M.,
  Polzik, E., and Sanpera, A.
\newblock Quantum polarization spectroscopy of correlations in attractive
  fermionic gases.
\newblock {\em New J. Phys.}{ \bf 11}(5), 055041 (2009).

\bibitem{bruun_probing_2009}
Bruun, G.~M., Andersen, B.~M., Demler, E., and {S\o rensen}, A.~S.
\newblock Probing spatial spin correlations of ultracold gases by quantum noise
  spectroscopy.
\newblock {\em Phys. Rev. Lett.}{ \bf 102}(3), 030401, January  (2009).

\bibitem{hammerer_quantum_2010}
Hammerer, K., {S\o rensen}, A.~S., and Polzik, E.~S.
\newblock Quantum interface between light and atomic ensembles.
\newblock {\em Rev. Mod. Phys.}{ \bf 82}(2), 1041, April  (2010).

\bibitem{esteve_observations_2006}
Est\`eve, J., Trebbia, J., Schumm, T., Aspect, A., Westbrook, C.~I., and
  Bouchoule, I.
\newblock Observations of density fluctuations in an elongated bose gas: Ideal
  gas and quasicondensate regimes.
\newblock {\em Phys. Rev. Lett.}{ \bf 96}(13), 130403, April  (2006).

\bibitem{gemelke_situ_2009}
Gemelke, N., Zhang, X., Hung, C., and Chin, C.
\newblock In situ observation of incompressible mott-insulating domains in
  ultracold atomic gases.
\newblock {\em Nature}{ \bf 460}(7258), 995--998 (2009).

\bibitem{muller_local_2010}
M\"uller, T., Zimmermann, B., Meineke, J., Brantut, J., Esslinger, T., and
  Moritz, H.
\newblock Local observation of antibunching in a trapped fermi gas.
\newblock {\em Phys. Rev. Lett.}{ \bf 105}(4), 040401, July  (2010).

\bibitem{sanner_suppression_2010}
Sanner, C., Su, E.~J., Keshet, A., Gommers, R., Shin, Y., Huang, W., and
  Ketterle, W.
\newblock Suppression of density fluctuations in a quantum degenerate fermi
  gas.
\newblock {\em Phys. Rev. Lett.}{ \bf 105}(4), 040402, July  (2010).

\bibitem{bakr_probing_2010}
Bakr, W.~S., Peng, A., Tai, M.~E., Ma, R., Simon, J., Gillen, J.~I.,
  F\"{o}lling, S., Pollet, L., and Greiner, M.
\newblock Probing the {Superfluid{\textendash}to{\textendash}Mott} insulator
  transition at the {Single-Atom} level.
\newblock {\em Science}{ \bf 329}(5991), 547 --550, July  (2010).

\bibitem{sherson_single-atom-resolved_2010}
Sherson, J.~F., Weitenberg, C., Endres, M., Cheneau, M., Bloch, I., and Kuhr,
  S.
\newblock Single-atom-resolved fluorescence imaging of an atomic mott
  insulator.
\newblock {\em Nature}{ \bf 467}(7311), 68--72 (2010).

\bibitem{serwane_deterministic_2011}
Serwane, F., {Z\"urn}, G., Lompe, T., Ottenstein, T.~B., Wenz, A.~N., and
  Jochim, S.
\newblock Deterministic preparation of a tunable {Few-Fermion} system.
\newblock {\em Science}{ \bf 332}(6027), 336 --338, April  (2011).

\bibitem{oblak_quantum-noise-limited_2005}
Oblak, D., Petrov, P.~G., Alzar, C. L.~G., Tittel, W., Vershovski, A.~K.,
  Mikkelsen, J.~K., {S\o rensen}, J.~L., and Polzik, E.~S.
\newblock Quantum-noise-limited interferometric measurement of atomic noise:
  Towards spin squeezing on the cs clock transition.
\newblock {\em Phys. Rev. A}{ \bf 71}(4), 043807, April  (2005).

\bibitem{appel_mesoscopic_2009}
Appel, J., Windpassinger, P.~J., Oblak, D., Hoff, U.~B., {Kj\ae rgaard}, N.,
  and Polzik, E.~S.
\newblock Mesoscopic atomic entanglement for precision measurements beyond the
  standard quantum limit.
\newblock {\em P. Natl. Acad. Sci. USA}{ \bf 106}(27), 10960 --10965, July
  (2009).

\bibitem{sanner_speckle_2011}
Sanner, C., Su, E.~J., Keshet, A., Huang, W., Gillen, J., Gommers, R., and
  Ketterle, W.
\newblock Speckle imaging of spin fluctuations in a strongly interacting fermi
  gas.
\newblock {\em Phys. Rev. Lett.}{ \bf 106}(1), 010402, January  (2011).

\bibitem{aljunid_phase_2009}
Aljunid, S.~A., Tey, M.~K., Chng, B., Liew, T., Maslennikov, G., Scarani, V.,
  and Kurtsiefer, C.
\newblock Phase shift of a weak coherent beam induced by a single atom.
\newblock {\em Phys. Rev. Lett.}{ \bf 103}(15), 153601, October  (2009).

\bibitem{lye_nondestructive_2003}
Lye, J.~E., Hope, J.~J., and Close, J.~D.
\newblock Nondestructive dynamic detectors for {Bose-Einstein} condensates.
\newblock {\em Phys. Rev. A}{ \bf 67}(4), 043609, April  (2003).

\bibitem{seo_compressibility_2011}
Seo, K. and {S\'{a} de Melo}, C. A.~R.
\newblock Compressibility and spin susceptibility in the evolution from {BCS}
  to {BEC} superfluids.
\newblock {\em Preprint at http://arXiv.org/abs/1105.4365}{ \bf
  \textcolor{white}{0}}, May  (2011).

\bibitem{klawunn_local_2011}
Klawunn, M., Recati, A., Pitaevskii, L.~P., and Stringari, S.
\newblock Local atom-number fluctuations in quantum gases at finite
  temperature.
\newblock {\em Phys. Rev. A}{ \bf 84}(3), 033612 (2011).

\bibitem{hung_observation_2011}
Hung, C., Zhang, X., Gemelke, N., and Chin, C.
\newblock Observation of scale invariance and universality in two-dimensional
  bose gases.
\newblock {\em Nature}{ \bf 470}(7333), 236--239, February  (2011).

\bibitem{wiesniak_magnetic_2005}
Wie\'sniak, M., Vedral, V., and \v{C}aslav Brukner.
\newblock Magnetic susceptibility as a macroscopic entanglement witness.
\newblock {\em New J. Phys.}{ \bf 7}, 258--258 (2005).

\bibitem{toth_spin_2009}
T\'{o}th, G., Knapp, C., G\"uhne, O., and Briegel, H.~J.
\newblock Spin squeezing and entanglement.
\newblock {\em Phys. Rev. A}{ \bf 79}(4), 042334, April  (2009).

\bibitem{Vedral2003}
Vedral, V.
\newblock Entanglement in the second quantization formalism.
\newblock {\em Central European Journal of Physics}{ \bf 1}(2), 289--306, June
  (2003).

\bibitem{Horodecki2009}
Horodecki, R., Horodecki, P., Horodecki, M., and Horodecki, K.
\newblock Quantum entanglement.
\newblock {\em Rev. Mod. Phys.}{ \bf 81}, 865--942, Jun  (2009).

\bibitem{kheruntsyan_quantum_2006}
Kheruntsyan, K.~V.
\newblock Quantum atom optics with fermions from molecular dissociation.
\newblock {\em Phys. Rev. Lett.}{ \bf 96}(11), 110401, Mar  (2006).

\bibitem{Zwierlein2003}
Zwierlein, M.~W., Hadzibabic, Z., Gupta, S., and Ketterle, W.
\newblock Spectroscopic insensitivity to cold collisions in a two-state mixture
  of fermions.
\newblock {\em Phys. Rev. Lett.}{ \bf 91}, 250404, Dec  (2003).

\bibitem{cherng_quantum_2007}
Cherng, R.~W. and Demler, E.
\newblock Quantum noise analysis of spin systems realized with cold atoms.
\newblock {\em New J. Phys.}{ \bf 9}(1), 7 (2007).

\bibitem{zimmermann_high-resolution_2011}
Zimmermann, B., M\"uller, T., Meineke, J., Esslinger, T., and Moritz, H.
\newblock High-resolution imaging of ultracold fermions in microscopically
  tailored optical potentials.
\newblock {\em New J. of Phys.}{ \bf 13}, 043007, April  (2011).

\bibitem{bradley_bose-einstein_1997}
Bradley, C.~C., Sackett, C.~A., and Hulet, R.~G.
\newblock {Bose-Einstein} condensation of lithium: Observation of limited
  condensate number.
\newblock {\em Phys. Rev. Lett.}{ \bf 78}(6), 985, February  (1997).

\bibitem{Huang_Statistical}
Huang, K.
\newblock {\em Statistical mechanics}.
\newblock Wiley, New York, 2nd ed. edition,  (1987).

\bibitem{giorgini_theory_2008}
Giorgini, S., Pitaevskii, L.~P., and Stringari, S.
\newblock Theory of ultracold atomic fermi gases.
\newblock {\em Rev. Mod. Phys.}{ \bf 80}(4), 1215--60 (2008).

\end{thebibliography}

\paragraph{Acknowledgments} We acknowledge enlightening discussions with M. Christiandl, A. Imamoglu, K. M\o lmer, E. Polzik, R. Renner, A. S\o rensen, M. Ueda and V. Vuletic and funding from NCCR MaNep, NCCR QSIT, ERC SQMS and FP7 FET-open NameQuam. JPB acknowledges support of European Union under Marie Curie IEF fellowship.

\paragraph{Author Contributions} JM and JPB analyzed the data, JM, JPB, DS and TM carried out the experimental work, all authors contributed to project planning and to writing the manuscript.

\paragraph{} Correspondence should be addressed to TE.

\paragraph{} The authors declare no competing financial interests.

\section{Beam generation}

\begin{figure}
  \includegraphics[width=0.45\textwidth]{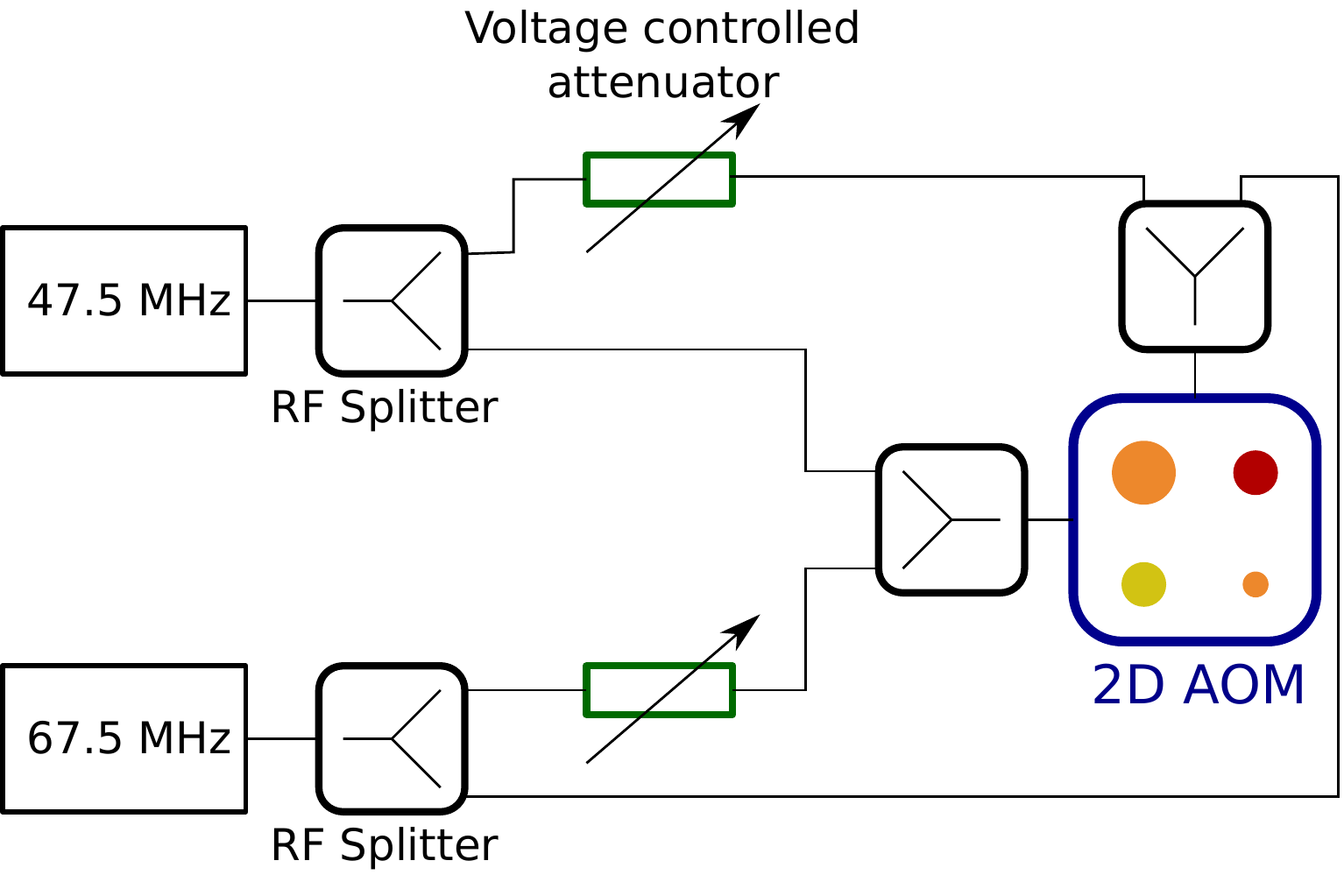}
  \caption{RF circuit for the generation of the interferometer beams. Two RF sources (USRP 2) generate signals at 47.5 and 67.5 MHz, respectively.  After splitting and selective attenuation, the two frequencies are combined and injected on each axis of the 2D AOM. Four beams are generated having a variable power ratio. The four spots in the image represent the four beams after passing through the microscope. The sizes of the spots represent the relative powers, and the colors represent different frequencies of the deflected light.} \label{figRF}
\end{figure}

The interferometer beams are generated using a two-axis acousto-optic deflector (2D AOD). Radio-frequency (RF) signals at two different frequencies are sent to both axes of the 2D AOD, resulting in the deflection of four beams in the -1/-1 order. Two RF sources (USRP 2) generate monochromatic signals, one at 47.5 MHz and the other at 67.5 MHz. The splitting and subsequent recombination depicted in Fig. \ref{figRF} allows to tune the power ratio between the two frequencies, for each axis of the 2D AOD.

After passing through a high-resolution microscope objective, the beams deflected in the -1/-1 order are arranged in a square and have a variable power, as depicted on Figure~\ref{figRF}. The most (least) powerful of those beams has been deflected by the non-attenuated (attenuated) signals from each source. These two beams have the same frequency, differing by 115 MHz from that of the incoming laser beam, and they lead to the interference pattern observed in the experiment. Conversely, the two beams having the same power have frequencies differing by 40 MHz, and contribute only to the background.

\section{Data processing using correlations}

\begin{figure*}
  \includegraphics[width=\textwidth]{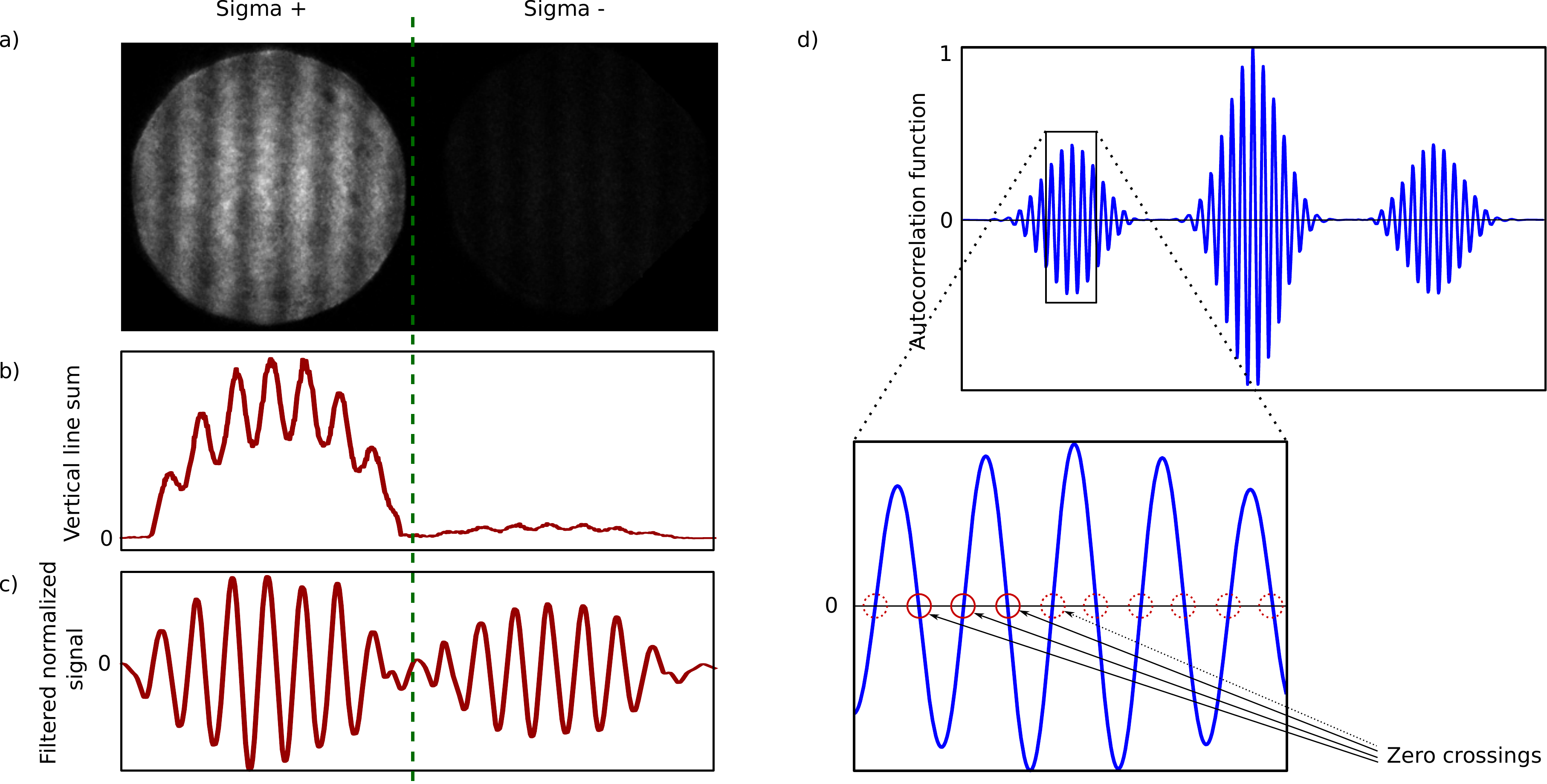}
  \caption{ Data processing using the correlation method. a) Raw picture observed on the camera, with the reference signal on the left ($\sigma^+$) and the probe signal on the right ($\sigma^-$). The power ratio of the two in 20. b) The signals from the picture is accumulated in the direction of the fringes, yielding a one-dimensional fringe pattern. The visibility of the fringes reflects the ratio of probe to local oscillator. c) The probe signal on the right side has been scaled up, and the full fringe pattern is spatially filtered leaving only the interference contribution. d) shows the autocorrelation function of the filtered fringe, where a fixed spacing between the two probe and the reference has been inserted. Thus, $\sigma^+$ - $\sigma^-$ correlations appears separated (at the both sides) from other contributions ($\sigma^+$ - $\sigma^+$ and $\sigma^-$ - $\sigma^-$ correlations). A region centered on the left side of the correlation signal, containing the $\sigma^+$ - $\sigma^-$ correlations, is extracted. In this window, the positions of the zero-crossings of the correlation function are measured (see text).} \label{suppfig1}
\end{figure*}

In order to take full advantage of the similarity between the probe signal ($\sigma^-$ polarized) and the reference signal ($\sigma^+$ polarized), we process the data using a method inspired by heterodyne detection. Figure~\ref{fig1} illustrates the different steps of this data processing algorithm. 

The two parts of the experimental images contain the probe and reference signals (Figure~\ref{suppfig1}a). A line sum along the direction of the fringes is first computed, yielding a one dimensional fringe pattern signal (Figure~\ref{suppfig1}b). The right part of the signal is scaled so that the probe and reference have roughly the same intensity. A filter in Fourier space is then applied to the full scaled signal, conserving only the Fourier components around the fringe spacing. The precise shape of the filter does not influence the obtained results, provided the low frequency components (containing the envelope of the two fringe patterns and the background) are removed. Figure~\ref{suppfig1}c presents a typical signal obtained after those processing steps.

We then compute the autocorrelation function of the processed fringe pattern. A fixed spacing is inserted between the probe and reference signal in the processed fringe pattern before the correlation function is computed. We do this so that when computing the autocorrelation function of the picture, the $\sigma^+$ - $\sigma^-$ correlation is isolated from the other contributions ($\sigma^+$ - $\sigma^+$ and $\sigma^-$ - $\sigma^-$ ) Figure~\ref{suppfig1}d presents a typical autocorrelation signal. The autocorrelation function displays two separated parts. The sum of the correlation functions of the reference with itself and the probe with itself appears at the center. Conversely, the correlation function of the probe with the reference appears at the sides. The center part of this probe-reference correlation signal is selected (in order to avoid finite size effects), and the positions of the zero crossings are extracted by linear interpolation of the discrete signal, as depicted on Figure~\ref{suppfig1}d. The error made by linear interpolation is typically two orders of magnitude lower than the photon shot-noise limit in our experiments. The position of a zero-crossing of the correlation function corresponds to the displacement of  the $\sigma^+$-pattern with respect to the $\sigma^-$-pattern that is required to overlap the maxima of one pattern with the zero-crossings of the second pattern. The determination of the zero-crossings of the correlation function is thus a direct measurement of the {\it relative displacement} of the $\sigma^+$-pattern with respect to the $\sigma^-$-pattern.

Each run of the experiment yields three pictures : one taken in the presence of the atoms, and two pictures taken in the absence of atoms (hereafter named pictures 1,2 and 3). The position of the crossings and thus the fringe displacements are measured on each of those pictures. The mean over the crossings yields the mean position of the fringe pattern in each picture.

To obtain the distribution of the fringe displacements, the experiments are repeated up to 400 times, over about 2 hours. Over this period, slow drifts of the phase on the "atoms" pictures, of up to 5 degrees, are observed, most probably due to temperature variations in the environment. In order to compensate those drifts, a sliding average (over 15 runs) of the positions of the crossings on picture 3 is computed. We take this as a measurement of the drift, and subtract this averaged signal from the positions measured on pictures 1 and 2. This subtraction operation amounts to measuring the displacement of each crossing of picture 1 to the corresponding crossing of picture 3, corrected for long term deviations. Taking the average of this corrected quantity over the crossings (i.e. averaging the position measured for all the pattern), we obtain the relative displacement of the fringes on pictures 1 and 2 compared to 3. The ratio of this mean displacement to the period of the fringes yields the phase shift observed on pictures 1 and 2 compared to 3.

We now have two phase shifts measured for each run of the experiment, with and without atoms. Thus, we obtain the statistical distribution of the phase shifts in the presence of the atoms from all the shifts of picture 1, together with the distribution of shifts on pictures 2, taken in exactly the same conditions but without atoms. To measure the phase variance due to the atoms, we subtract the phase variance on picture 2 (background variance in the text) to the phase variance of the picture 1. 

\section{Entanglement Witness}
In reference \cite{wiesniak_magnetic_2005} it is shown that the magnetic susceptibility is a macroscopic entanglement witness. If the magnetic susceptibility is lower than a certain bound, the state describing the thermal state of the bulk system cannot be a convex combination of product states (separable state), which means that the state must contain entanglement.
The arguments in ref. \cite{wiesniak_magnetic_2005} are based on the fluctuation-dissipation theorem which links the magnetic susceptibility to the fluctuations of the spins that are the constituents of the system. The experimental signal is proportional to the difference in atomic density in each of the two spin states. Let us define the collective spin describing the magnetization as
\begin{equation}
 \vec{M}=\sum \vec{\sigma}_i,
\end{equation}
where $\vec{\sigma}_i$ are Pauli matrices. The calculation described in ref. \cite{wiesniak_magnetic_2005} leads to
\begin{equation}
 \delta M_x^2+\delta M_y^2+\delta M_z^2\geq 2 N. \label{eq:ineq}
\end{equation}
Assuming that the fluctuations are the same along all three axes, as we will argue below, the fluctuations in a given direction are bounded by
\begin{equation}\label{eq:rotsymineq}
 \delta M_z^2\geq \frac{2}{3} N.
\end{equation}
In the experiment we have access to the column-integrated total density $n_{\rm{col}}=N/A$ and the column-integrated fluctuation density $\delta m^2=\delta M_z^2/A^2$, where $A$ is the effective area of the probed region. The effective area $A$ is determined experimentally and takes into account the intensity profile of the probe beam. Using this in \eqref{eq:rotsymineq} we get
\begin{equation}
 \frac{A\delta m^2}{n_{\rm{col}}}\geq \frac{2}{3}.
\end{equation}
The measurement of a value lower than $2/3$ means that the state cannot be separable.

\subsection*{Invariance under Spin Rotations}
The Hamiltonian describing a two-component Fermi gas is given by \cite{giorgini_theory_2008}:
\begin{eqnarray}
 H&=&\sum_{\sigma}\int d\vec{r}\Psi(\vec{r})_{\sigma}^{\dagger}\left(-\frac{\hbar^2 \nabla^2}{2m}+V_{ext}(\vec{r})-\mu_{\sigma}\right)\Psi(\vec{r})_{\sigma}\nonumber \\
&+	&\iint d\vec{r}d\vec{r'}V(\vec{r}-\vec{r'})\Psi(\vec{r})_{\uparrow}^{\dagger}\Psi(\vec{r'})_{\downarrow}^{\dagger}\Psi(\vec{r'})_{\downarrow}\Psi(\vec{r})_{\uparrow}.
\end{eqnarray}
Here the field operators for the two spin states denoted by $\sigma\in\{\uparrow, \downarrow\}$ obey the fermionic anti-commutation relation $\{\Psi(\vec{r})_{\sigma},\Psi(\vec{r'})_{\sigma'}^{\dagger}\}=\delta_{\sigma,\sigma'}\delta(\vec{r}-\vec{r'})$, $V$ is the interaction potential, $V_{ext}$ the trapping potential and $\mu_{\sigma}$ is the chemical potential for the two components.
At the energy scale of our experiment, the interaction term in the Hamiltonian describes the s-wave scattering of two particles, the strength of which is characterized by the scattering length $a$. The wavefunction describing the relative position of the two particles is symmetric under the exchange of particles. As a result the spin wavefunction must be antisymmetric (singlet) and is thus invariant under rotations of the spin.
For a balanced gas, $\mu_{\uparrow}=\mu_{\downarrow}$, the first part of the Hamiltonian is invariant under rotations of the spin state. The external trapping potential for the two components differs by the energy of the hyperfine splitting. However, since no transitions between the states occur on experimentally relevant timescales and since we prepare a balanced gas, this energy difference is irrelevant for the dynamics of the system.
Since we prepare the system in a thermal state, the transverse components of the collective spin $\mathbf{M}$ have completely decohered. The invariance under rotations of the spin relies both on the structure of the Hamiltonian and on the preparation of a balanced, thermal state~\cite{Zwierlein2003}

\subsection*{Entanglement Witness in Grand-Canonical Ensemble}

\paragraph{Symmetry for grand-canonical expectation values}
In our experiment we measure the spin-fluctuations of a small subsystem of the whole atomic cloud. This subsystem is in thermodynamic equilibrium with the rest of the system. In this situation the expectation values in the grand-canonical ensemble of a given observable have the same symmetry as the Hamiltonian, which might not be the case for the ground state. For the specific case considered we have $\langle M_z\rangle=\langle U M_z U^{\dagger}\rangle$, where $U$ is an arbitrary rotation of the spin state. The state of the system is given by the density matrix $\rho=\frac{1}{Z}e^{-\beta H}$. We then have
\begin{eqnarray}
 \langle U M_z U^{\dagger}\rangle&=& Tr \rho U M_z U^{\dagger}\nonumber\\
  &=& \frac{1}{Z}Tre^{-\beta H}U M_z U^{\dagger} \nonumber\\
  &=& \frac{1}{Z}Tr Ue^{-\beta H} M_z U^{\dagger}\quad([H,U]=0) \nonumber\\
  &=& \frac{1}{Z}Tr e^{-\beta H} M_z U^{\dagger}U\quad(\mathrm{trace\ cyclic}) \nonumber\\
  &=& \frac{1}{Z}Tr e^{-\beta H} M_z \nonumber\\
  &=& \langle M_z \rangle.
\end{eqnarray}
As an example, consider the situation of a ferromagnetic ground state where domains of spins pointing in a certain direction would exist. In each realization the experiment the symmetry is broken, but the direction in which the spins are pointing is different. Every direction must be equally probable, which leads to the situation that the ensemble averages have the same symmetry than the Hamiltonian.
\paragraph{Non-fixed particle number}
The derivation of the inequalitiy \ref{eq:ineq} assumes a fixed number of particles $N$ and the authors of ref. \cite{wiesniak_magnetic_2005} work in the canonical ensemble. However, the argument carries through for measurements in the grand-canonical ensemble where $N$ is replaced by its expectation value $\langle N\rangle$. For this we express the density matrix in the grand-canonical ensemble as
\begin{equation}
\rho=\sum_N w_N \rho_N,
\end{equation}
where $\rho_N$ is the density matrix in the subspace of fixed particle number $N$ with the weight factors $w_N$. As before, we write $\langle \cdot \rangle=Tr \rho \cdot$ for the grand-canonical expectation value and similarly we define $\langle \cdot \rangle_N=Tr \rho_N \cdot$ as the expectation value in the $N$-particle subspace. Similarly, $\delta\cdot$ and $\delta_N \cdot$ denote the variance in the grand-canonical ensemble and with fixed particle number $N$, respectively. We now calculate the variance of the operator $M_z$
\begin{eqnarray}
 \delta M_z^2&=&\langle (M_z -\langle M_z \rangle)^2\rangle\nonumber\\
&=&\sum_N w_N Tr \rho_N (M_z -\langle M_z \rangle)^2\nonumber\\
&=&\sum_N w_N Tr \rho_N (M_z -\langle M_z \rangle_N -(\langle M_z \rangle-\langle M_z \rangle_N))^2\nonumber\\
&=&\sum_N w_N Tr \rho_N \Big{[}(M_z -\langle M_z \rangle_N)^2 \nonumber\\
&&-2( M_z -\langle M_z \rangle_N)(\langle M_z \rangle-\langle M_z \rangle_N)\nonumber\\
&&+(\langle M_z \rangle-\langle M_z \rangle_N)^2\Big{]}\nonumber\\
&=&\sum_N w_N \delta_N M_z^2 + \sum_N w_N(\langle M_z \rangle -\langle M_z \rangle_N)^2\nonumber\\
&\geq&\frac{2}{3}\langle N \rangle +\sum_N w_N(\langle M_z \rangle -\langle M_z \rangle_N)^2.
\end{eqnarray}
In the last step we have used the known result from \ref{eq:rotsymineq}. The last term in the last line is small, because the width of the weights $w_N$ is centered around $N=\langle N \rangle$ with a width $\sqrt{\langle N\rangle}$ and because $\langle M_z\rangle \simeq \langle M_z\rangle_{\langle N\rangle}$. More importantly, the last term is positive and if the fluctuations per particle are measured to be smaller than $2/3$ the state cannot be separable.

\end{document}